\documentclass[aps,prl,twocolumn,groupedaddress,showpacs]{revtex4}
\usepackage{graphicx}
\graphicspath{{figure file fold path}}
\usepackage[dvipsnames,usenames]{color}
\usepackage{color}
\usepackage{amsmath}
\usepackage{amssymb}

\emergencystretch=\maxdimen
\begin{document}
\title{Exploration of Micro-VCSEL Ultra-low Biasing Scheme for Pulse Generation}
\author{T.~Wang$^{1,2}$, G.F. Wang $^{1}$, G.P.~Puccioni$^{3}$, G.L.~Lippi,$^{2}$}

\affiliation{$^1$College of Electronics and Information, Hangzhou Dianzi University, Hangzhou 310018, China}
\affiliation{$^2$Universit\'e C\^ote d'Azur, Institut de Physique de Nice (INPHYNI), CNRS UMR 7010, Nice, France}
\affiliation{$^3$Istituto dei Sistemi Complessi, CNR, Via Madonna del Piano 10, I-50019 Sesto Fiorentino, Italy}

\begin{abstract}
The generation of optical pulses at ultra low bias level, thus low energy cost, is explored in a commercial microcavity semiconductor laser in view of testing the principle of energy efficient
information encoding in potential integrated schemes. Sequences of regular, highly nonlinear pulses with acceptable amplitude stability are obtained from a commercial device as potential sources of
bits where the information is added by post-treatment (pulse removal). A discussion on the energy expenditure per bit is offered, together with the optimal frequency for pulse generation, which is
found to lie slightly below the above-threshold value declared by the manufacturer. 
\end{abstract}

\pacs{}
\maketitle
%

\section{Introduction}
The great success in laser miniaturization has opened up a clear route towards the realization of integrated light sources at chip level ~\cite{Song2005, Khajavikhan2012, Hill2014} on the way towards the realization of all-optical chips ~\cite{Smit2012, Soref2018, Norman2018}. Very small lasers are considered promising candidates for optical interconnects ~\cite{Service2010, Chen2011, Huang2014} for communications and information processing, in addition to permitting advances in biosensing, medical imaging and spectroscopy ~\cite{Willets2007, Anker2008, Gourley2008, Kim2013, Lozovik2014, Abe2015}. One of the objectives of the miniaturization is the production of bits of information generated with a minimum amount of power and with minimal thermal load for the device and, ultimately, for the chip. Nanodevices may be a good solution to this quest as long as they are proven capable of generating usable (i.e., sufficiently coherent) optical pulses at very low input power levels.

\begin{figure}[!t]
\centering
\includegraphics[width=0.95\linewidth,clip=true]{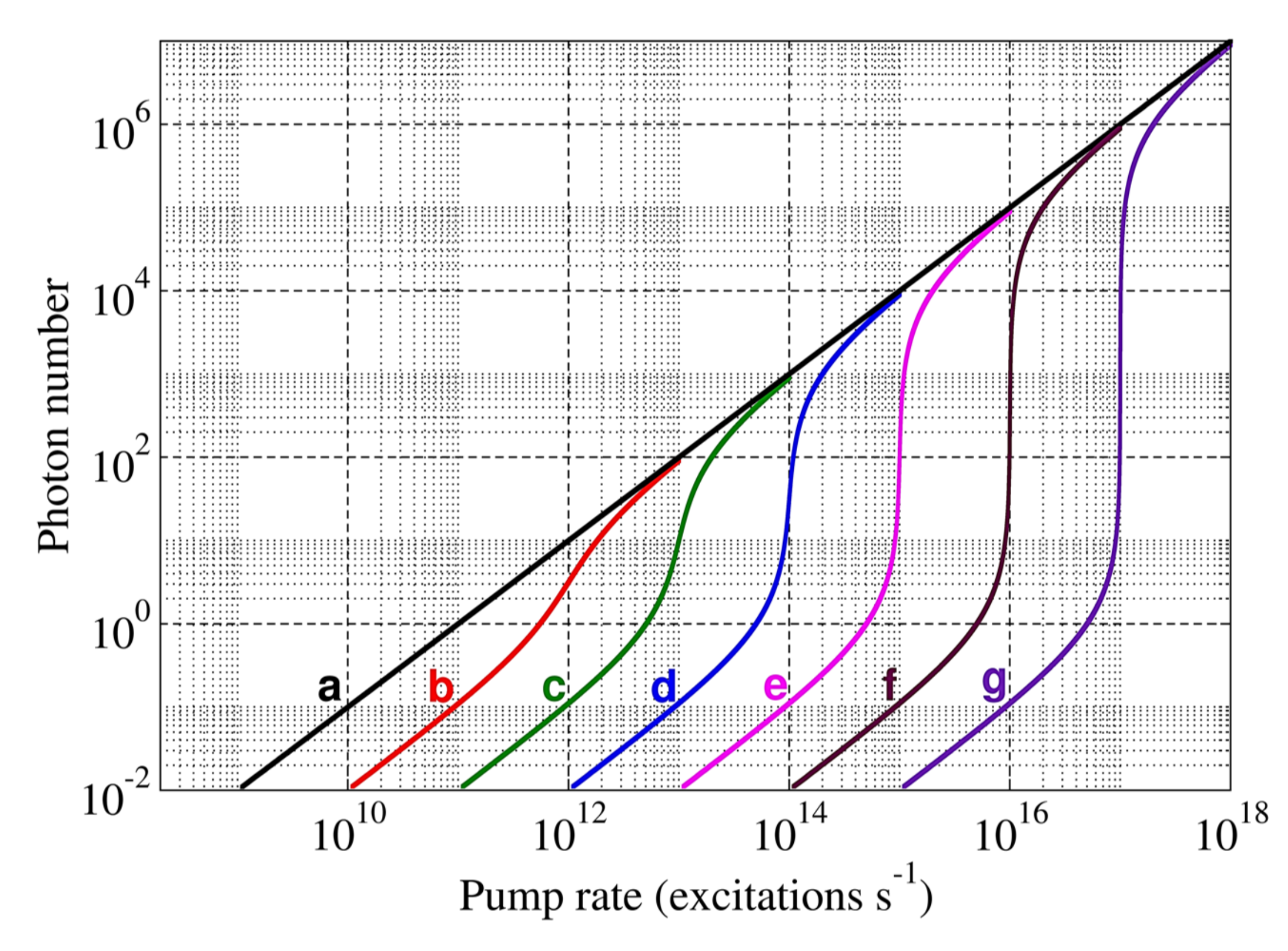}
\caption{Laser output (coherent photon number) as a function of input power (measured in number of excitations per second) in lasers with different fraction of spontaneous emission coupled into the lasing mode. $\beta$: 1 (a, black), $10^{-1}$ (b, red), $10^{-2}$ (c, green), $10^{-3}$ (d, blue), $10^{-4}$ (e, magenta), $10^{-5}$ (f, maroon), $10^{-6}$ (g, indigo). On the basis of considerations on the fluctuations ~\cite{Rice1994}, the devices are considered to be macroscopic ($\beta \lesssim 10^{-5}$ ), mesoscopic ($10^{-4} \lesssim \beta \lesssim 10^{-2}$) or nanoscopic ($10^{-1} \lesssim \beta \lesssim 1$) ~\cite{Wang2018}. As an example, one can compare the emission for $10^{13}$ excitations per second: for $\beta = 10^{-1} , 10^{-2} , 10^{-3} , 10^{-4}$, the average number of emitted
photons is $\langle n \rangle = 10^2 , 10^1 , 10^{-1} , 10^{-2}$, respectively.}
\label{scurves}
\end{figure}

The reduction in cavity volume is accompanied by a proportionally larger (coherent) photon flux at low excitation (cf. Fig. \ref{scurves}): at $10^{13}$ excitations per second devices with $\beta$ = 0.1 (fraction of spontaneous emission coupled into the lasing mode) have already switched to the upper branch of emission, while larger devices (smaller $\beta$ values) are still emitting only spontaneous photons, thus not only incoherent radiation, but also with a strongly reduced flux (details in figure caption). In order to exploit the advantage of lower energy expenditure for communications, two conditions need to be obviously fulfilled: 1. the emission needs to be sufficiently coherent, and 2. the pulse generation needs to be obtained as close to ``threshold'' as possible. These two conditions are not trivial since the identification of coherence is still a point which raises questions, at least for most practical devices, and the threshold definition is still contentious ~\cite{Editorial2017}.

It is obvious that raising the excitation rate - regardless of any other considerations - will always provide coherent output and a high-quality signal for information encoding, even from a nanolaser. This strategy, however, does not take into account the possible damage, or reduced lifetime, which may ensue for the device ~\cite{Khurgin2012}, nor does it fulfill the goal of minimizing power expenditure and thermal load. Thus, it is strategically important to explore the possibility of exploiting the smoother turn-on of smaller (i.e., larger $\beta$) lasers for data encoding.

Given the difficulties in detection posed by nanolasers, for this investigation we use a mesoscale device as surrogate ~\cite{Wang2018} (a micro-Vertical Cavity Surface Emitting Laser - VCSEL - corresponding approximately to the magenta line, curve e, in Fig. \ref{scurves}) and investigate the possibility of generating usable pulses in the nonlinear amplification region of the laser response (i.e., the steeper part of the curve). This approach does not answer the questions related to the field coherence, since our microVCSEL has a sufficiently large cavity volume to ensure coherence in the emission ~\cite{Wang2015}, but will test the potential for low energy-consumption optical pulses. 

\section{Experiment and analysis of observations}
\begin{figure}[!t]
\centering
\includegraphics[width=0.95\linewidth,clip=true]{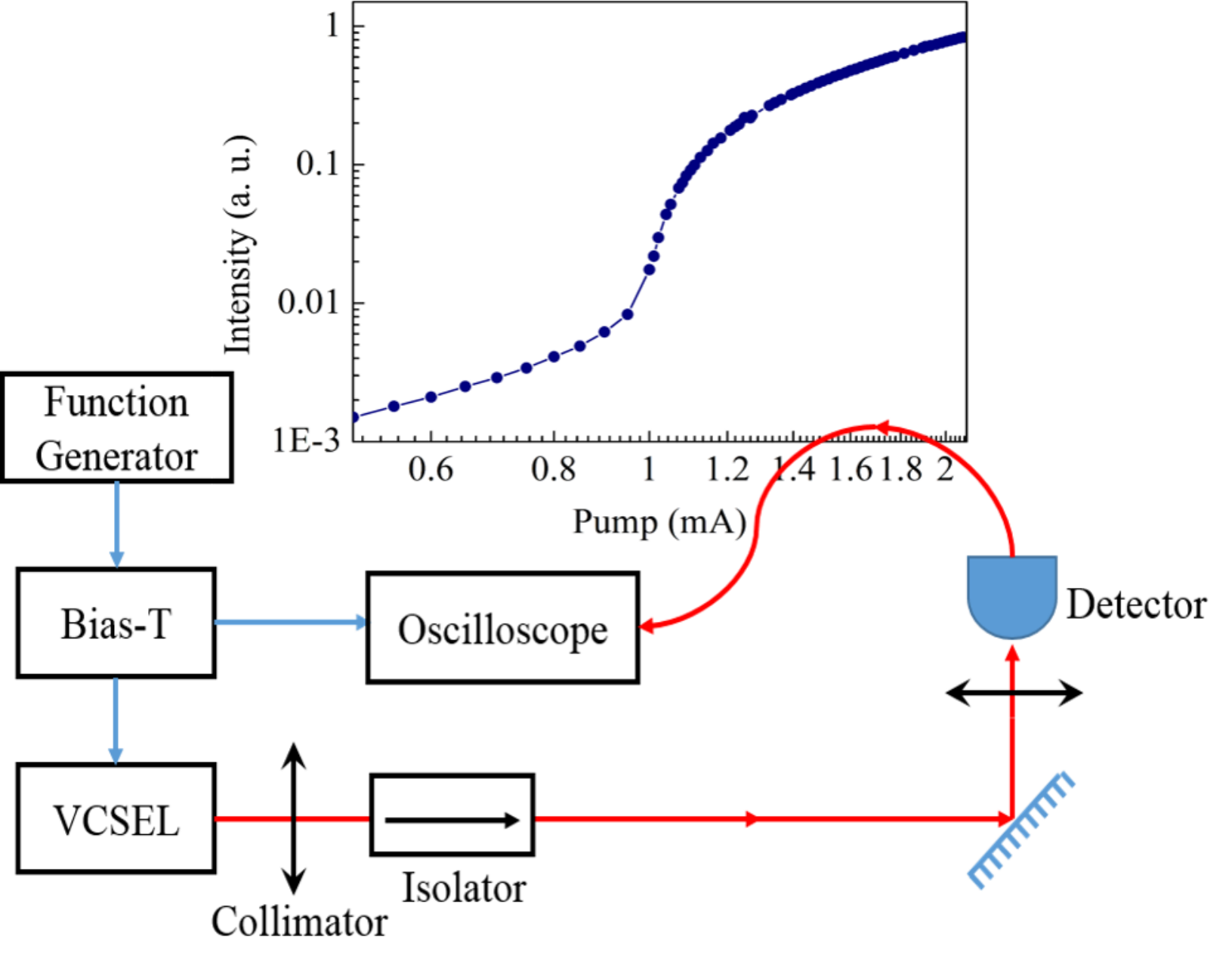}
\caption{Experimental setup. The laser is mounted on a TEC module (Thorlabs TCLDM9) which allows for temperature stabilization from an external source. We set the working temperature at $25^{\circ}$C. The current is supplied by a commercial dc power supply (Thorlabs LDC200VCSEL) with resolution $1\mu A$ and accuracy $\pm 20 \mu A$. The current drift at constant ambient temperature is less than $1\mu A$ over 30 minutes and both noise and ripple are smaller than $1.5 \mu A$. Additional details in text. The upper part of the figure (dark blue curve with dots) represents the input-output characteristic measured for this device.}
  \label{ExperiIV}
\end{figure}

The experimental setup is shown in Fig. \ref{ExperiIV}. The VCSEL (Thorlabs980) is designed for data transmission at $f_{dt} = 2.5 GHz$, has a $\beta \approx 10^{-4}$ ~\cite{Wang2015} with nominal emission wavelength $\lambda = 980 nm$, nominal threshold current $2.2 mA \lessapprox i_{th} \lessapprox 3.0 mA$ and maximum output power $P_{max} = 1.85 mW$ ~\cite{Manufac}. The laser is temperature -- stabilized to better than $0.1^{\circ}$C in a TEC module (Thorlabs TCLDM9). The laser output is collimated and passed through a Faraday isolator (QIOPTIQ8450-103-600-4-FI-980-SC) which prevents feedback into the laser. The optical signal is sampled by a fast, amplified photodetector (Thorlabs PDA8GS) with $9.5 GHz$ bandwidth and the signal is digitized by a LeCroy Wave Master 8600A with $6 GHz$ analog bandwidth and up to $5 \times 10^6$ sampled points per trace. The data are stored in a computer through a GPIB interface using Python. The sinusoidal modulation, introduced through a bias-tee, is provided by a function generator (E4421B, HEWLETT) at the chosen frequency $f_{mod} = 1 GHz$ to produce a modulation amplitude in the laser $i_{mod} \approx 0.3 mA$. The modulation amplitude is chosen in such a way as to obtain the most effective pulse generation with minimal amplitude (cf. below for evidence and ~\cite{Wang2018a} for spectral analysis). The choice of a sinusoidal modulation originates from its spectral purity, which allows for reliable results without the need for special probes. It can of course be extended to other forms, e.g. square modulation, for further investigations.

We first present the laser response for a small set of biasing choices, identifying the optimum condition, then analyze it. Thanks to previous work ~\cite{Wang2018a}, which allows us to identify the threshold for coherent emission, we know for this device that the latter is placed at $i_{b} = 1.06 mA$. The laser response (input-output or L-L curve) is displayed in the upper part of Fig. \ref{ExperiIV}, where the dots represent the measurements and the line is inserted to guide the eye. Notice that the manufacturer's declared ``threshold'' is beyond the right edge of the graph, thus in a region well distinct from the one that we investigate.

\begin{figure}[!t]
\centering
\includegraphics[width=0.95\linewidth,clip=true]{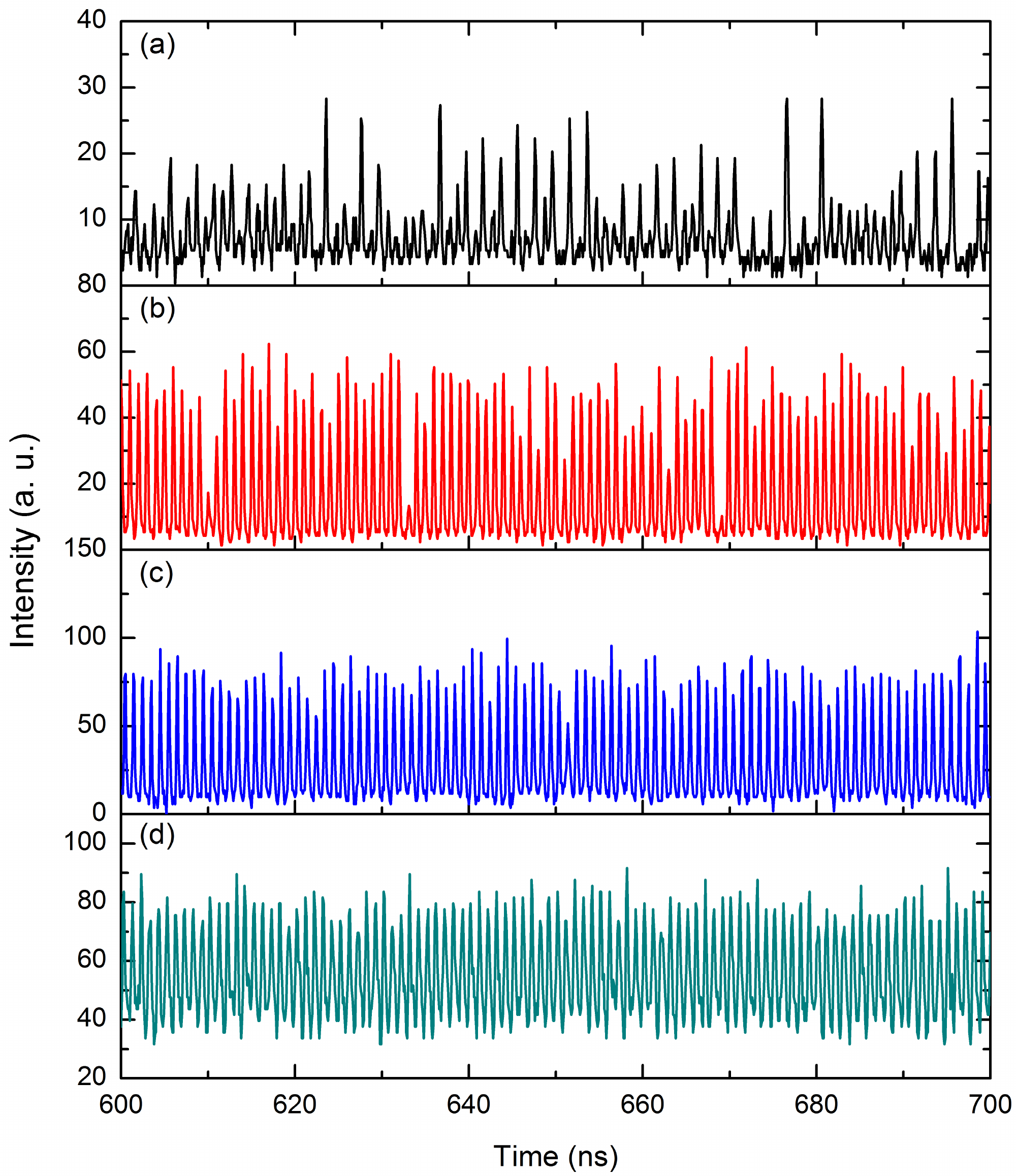}
  \caption{Typical dynamics within $100 ns$ of laser under modulation at $I_{b} = 0.95 mA$ (a); $1.10 mA$ (b); $1.15 mA$ (c);
$1.40 mA$ (d). Notice the change in vertical scale from each figure to the next for display optimization.}
  \label{dynamics}
\end{figure}

Fig.\ref{dynamics} shows the representative dynamics when the laser is modulated around the bias points: $i_{b} = 0.95mA$,
$1.10 mA$, $1.15 mA$, and $1.40 mA$. At $i_{b} = 0.95mA$, as shown in Fig.\ref{dynamics} (a), random spikes with variable amplitude are observed. This bias choice corresponds to the lowest point in the nonlinear branch in the output power which connects the spontaneous emission and the stimulated emission branches (L-L curve in Fig. \ref{ExperiIV}). This implies that the laser spends half of the modulation period in the spontaneous emission regime and is periodically driven up the amplification region, but only about two-thirds of the way up the nonlinear curve. Thus, the amplification process leading to lasing starts every time from a different initial condition and undergoes the typical delay at turn-on ~\cite{Porta1998, Porta2000} common to all Class B lasers ~\cite{Arecchi1984,Tredicce1985}, with a degree of variation in the turn-on time which depends on the actual value of the spontaneous contribution to the lasing mode changing at each cycle. This is the reason for the strong differences in success rate for generating a pulse at any given modulation period. As soon as the bias current is sufficiently large -- $i_{b} = 1.10mA$, thus $\approx 4 \%$ above the measured threshold for coherent oscillation ~\cite{Wang2018a} - a sequence of regular pulses with unequal amplitudes is generated, as shown in Fig.\ref{dynamics}(b). For this bias choice, the laser is regularly brought below the kink-current value, but only for approximately one third of the modulation period and therefore not as deeply into the spontaneous emission regime. As a consequence, the pulses become rather regular: one laser output pulse is generated for each modulation period, albeit with widely varying amplitudes. The reason is the same as observed for the lower biasing, but the influence of the initial condition (contribution of the on-axis spontaneous emission at the turn-on instant) now is reflected only onto the peak amplitude with little consequence for the time delay at turn-on and none (at least on reasonably long time sequences) for the success rate in pulse generation.

Increasing slightly the bias point, $i_{b} = 1.15 mA$ (i.e., $\approx 8 \%$ above threshold ~\cite{Wang2018a}), we end up with an optimal condition, where we can consistently obtain regular pulses which, while displaying some residual variability, can be well identified with the help of thresholding: setting a threshold at, for instance, $50 \%$ of the average peak level we can confirm the presence of a pulse per modulation period. During the modulation, the laser is still brought temporarily below threshold, but only for a time $\tau_{bt} \approx 0.25 ns$. Once the pump is below threshold, the relaxation occurs on the carriers' time scale $\gamma^{-1} \approx 0.3 ns$, which is substantially of the same order, but actually slightly longer than, $\tau_{bt}$. As a consequence, the laser dips down below threshold, turning off almost entirely (cf. Fig. \ref{dynamics}c), but the population inversion does not reach its below threshold state before restarting. This enables the establishment of a memory of the state which preceded the turn-off of the photon field and the laser can restart without suffering almost any of the ``ill effects'' coming from the randomness of the spontaneous emission. The result is a regular series of pulses whose amplitude does not vary excessively.

\begin{figure}[!t]
\centering
  \includegraphics[width=0.95\linewidth,clip=true]{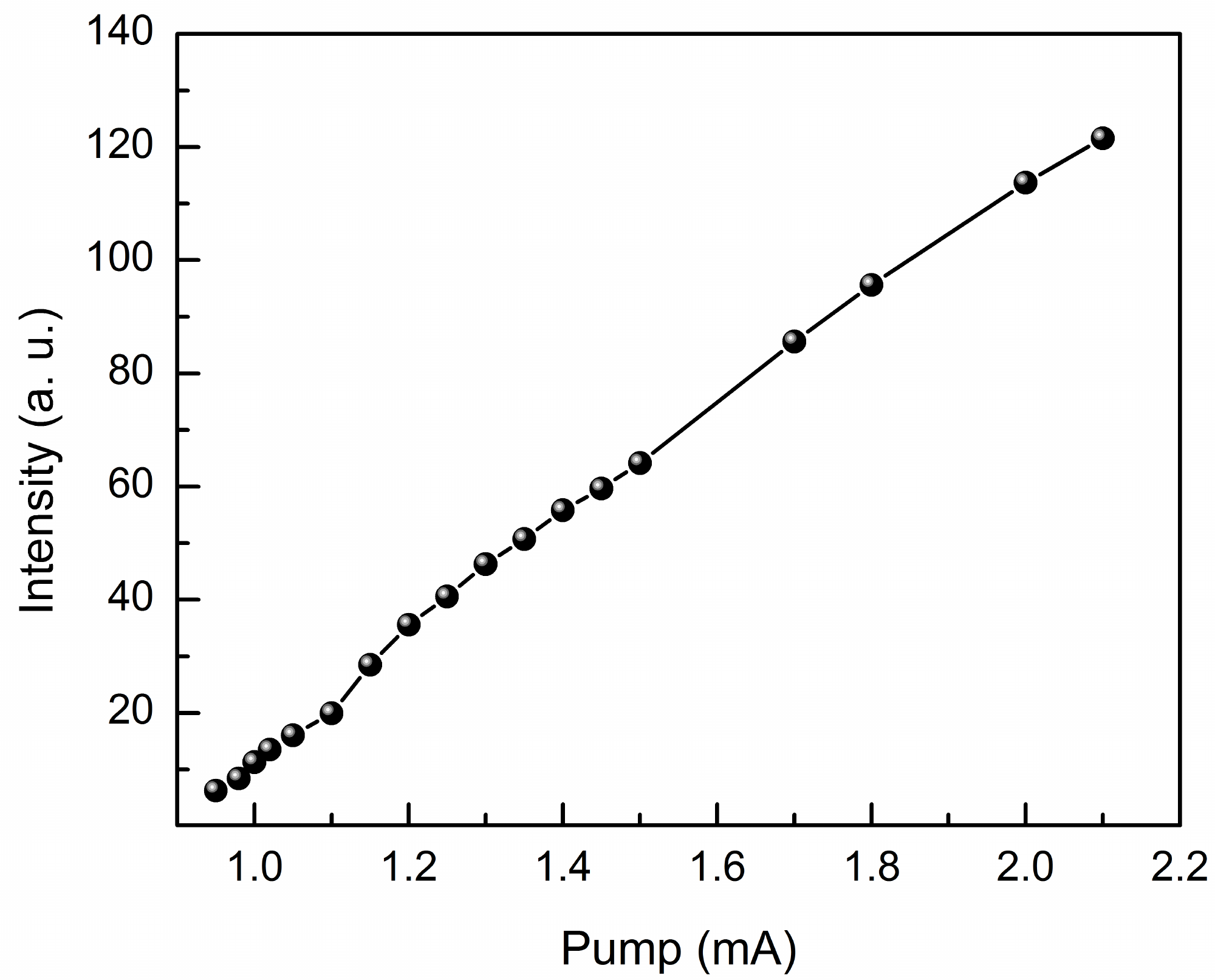}
  \caption{Average intensity emitted by the laser, in the presence of modulation, in a pump interval which covers the transition
between the lower and the upper branch. The scales here are linear and the response nearly linear, since we are looking at a detail of the L-L curve (Fig. \ref{ExperiIV}).}
  \label{IVsingle}
\end{figure}

In order to get an idea for the amount of optical output emitted, we estimate the cw output power at $P_{cw} (i_{b} =
1.2mA) \approx 50 \mu W$ , while the peak power for Fig. \ref{dynamics}c is ($120 \lessapprox P_{max} \lessapprox 160$) $\mu W$. Thus, the pulsed operation provides a peak output level which is approximately three times the cw equivalent. It is interesting to notice, however, that even at the level of average output, the modulation raises somewhat the average values: Fig. \ref{IVsingle} shows a small bump in the average output in correspondence to the bias pump interval where the modulation is most effective (Fig. \ref{dynamics}).However, its average contribution is of the order of 8 $\mu W$ (or $< 20$\%), thus still very small compared to the factor 3 gained at the peak.

Finally, by raising the bias to the top of the nonlinear amplification regime connecting the two main branches of laser output ($i_{b} = 1.4mA$), we see that the laser never turns off for our choice of modulation amplitude (Fig.\ref{dynamics}d). The modulation now prevents the system from turning off the photon field and, rather than obtaining pulses, we end up with regular oscillations, albeit somewhat noisy.

The reason for our choice of modulation amplitude is that, as previously explained, we are searching for conditions which allow for the generation of well-defined pulses of minimal amplitude. If we had chosen a smaller amplitude, the system would have still been too influenced by noise, since the biasing point fulfilling the condition of Fig. \ref{dynamics}c would have been a lot further down (rather than approximately half-way up the nonlinear amplification region). If we had chosen a larger amplitude, we would have obtained larger pulses and would have entered the fully lasing region (upper branch), thus emitting a larger amount of power but also requiring more current and larger modulation. Thus, the choice of modulation amplitude and bias point is a compromise between the two requirements, where the relationship between the two choices is dictated by the condition explicited above for preserving the memory of the previous state in the carrier variable: $\tau_{bt} \approx \gamma^{-1}$.

\begin{figure}[!t]
\centering
  \includegraphics[width=0.95\linewidth,clip=true]{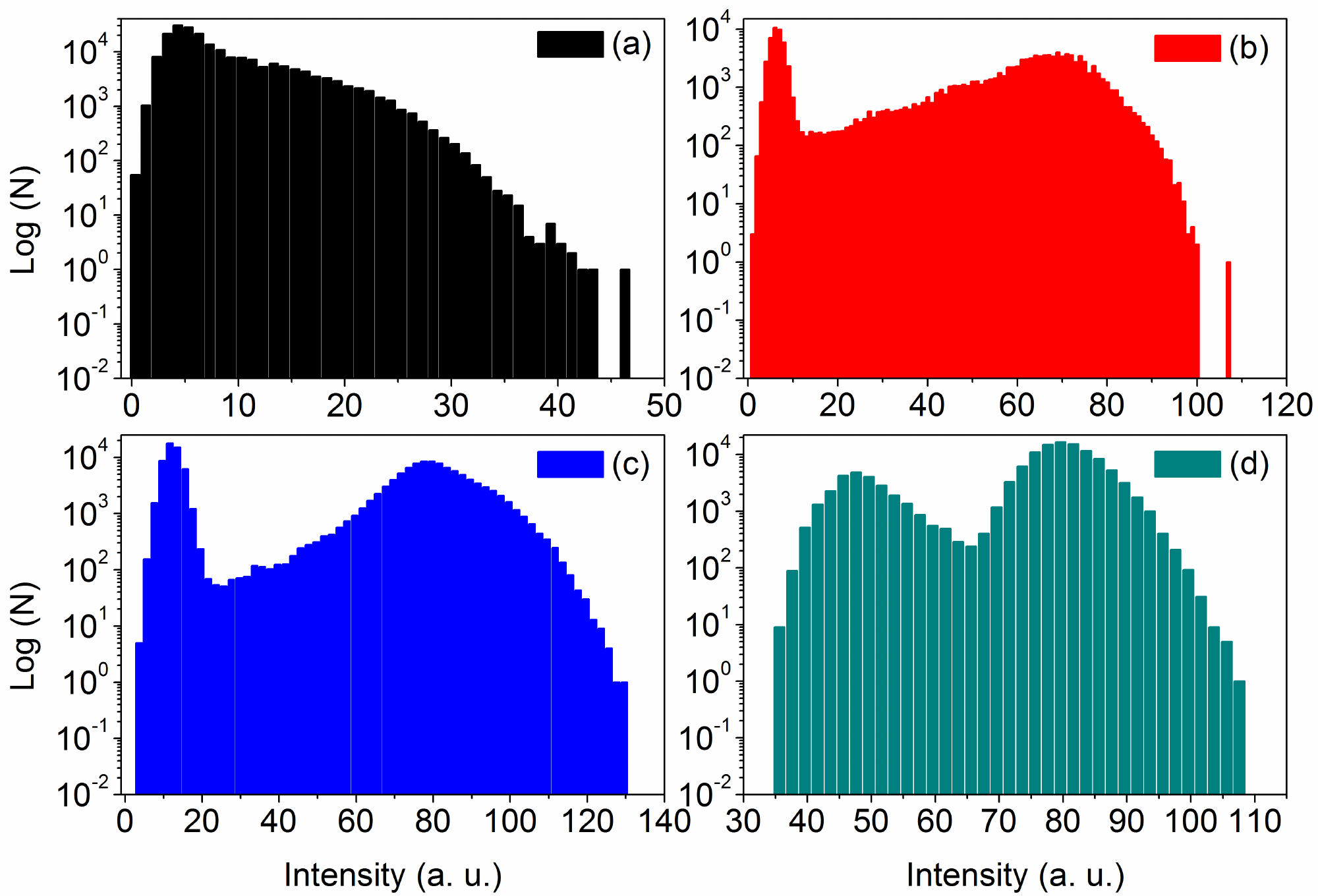}
  \caption{Intensity distribution histograms of laser response for the bias current at: (a) $0.95 mA$; (b) $1.10 mA$; (c) $1.15mA$; (d) $1.40 mA$.}
  \label{propability}
\end{figure}

A complementary characterization of the modulation results can be obtained by plotting the histograms of the temporal traces. Fig. \ref{propability} shows the histograms of the mesured intensity values in semilogarithmic scale. Panel (a) corresponds to the lowest biasing ($i_{b} = 0.95mA$) and clearly shows the lack of any structure in the data: the frequency of occurrence for each intensity value decreases monotonically. A structure at high intensity develops in panel (b) ($i_{b} = 1.10mA$), but the transition from the left peak - ``zero'' intensity, i.e., pulse background -- and the peak is not so clearly defined since the growth is rather gradual. The distinction between background and peak intensity is quite clear in panel (c) ($i_{b} = 1.15mA$), confirming the result of Fig.\ref{dynamics}c which showed pulses of fairly regular amplitude. Notice that the tail towards high peaks corresponds to quite rare occurrences (more than three orders of magnitude between the maximum of the distribution and the maximum intensity). The most populated intensity level is found around 80 and good discrimination for the pulse occurrence can be placed around 50, in agreement with what observable from Fig.\ref{dynamics}c. Finally, for large bias ($i_{b} = 1.4mA$) the two peaks in the histogram correspond to the lowest and largest value of the intensity during the noisy oscillation (panel (d)). As already seen in the
temporal trace, this biasing choice does not offer a viable way of encoding information due to its low contrast and ill-defined levels.

\begin{figure}[!t]
\centering
  \includegraphics[width=0.95\linewidth,clip=true]{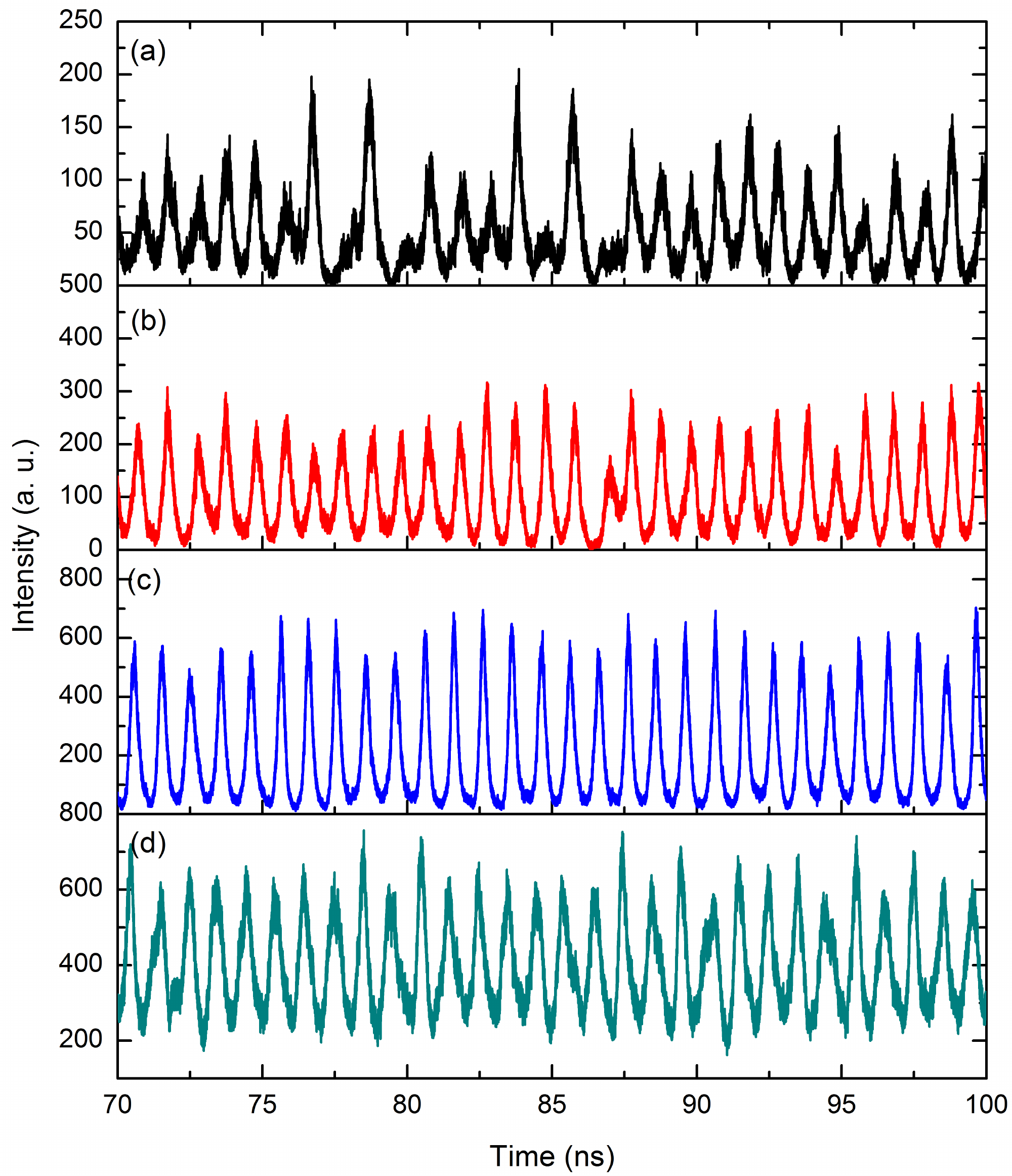}
  \caption{Calculated dynamics of laser modulated at $P = (a) 0.95P_{th}$; (b) $1.10P_{th}$ ; (c) $1.20P_{th}$ ; (d) $1.40P_{th}$. The modulation amplitude is $0.3P_{th}$ and the frequency is, as in the experiment, $1GHz$. The parameter values for the simulation are those already used for this laser ~\cite{Wang2015}.}
  \label{S-dynamics}
\end{figure}

As a final step, we can run a numerical simulation under the same conditions as the experiment. In order to properly account for the below threshold dynamics, governed by spontaneous emission, we use a recently developed Stochastic Simulator ~\cite{Puccioni2015} capable of fully accounting for these processes within the semiclassical description of the radiation-matter interaction ~\cite{Einstein1917}. The predictions, computed for four values of pump matching the experimental current values, are shown in Fig.\ref{S-dynamics} for a $\beta = 10^{-4}$ laser. The temporal dynamics shows strong similarities with the one observed in the experiment, thus confirming that the observed dynamics is the result of the dynamical pump modulation in the transition region.

\section{Energy considerations and scheme's potential}
\begin{figure}[!t]
\centering
  \includegraphics[width=0.95\linewidth,clip=true]{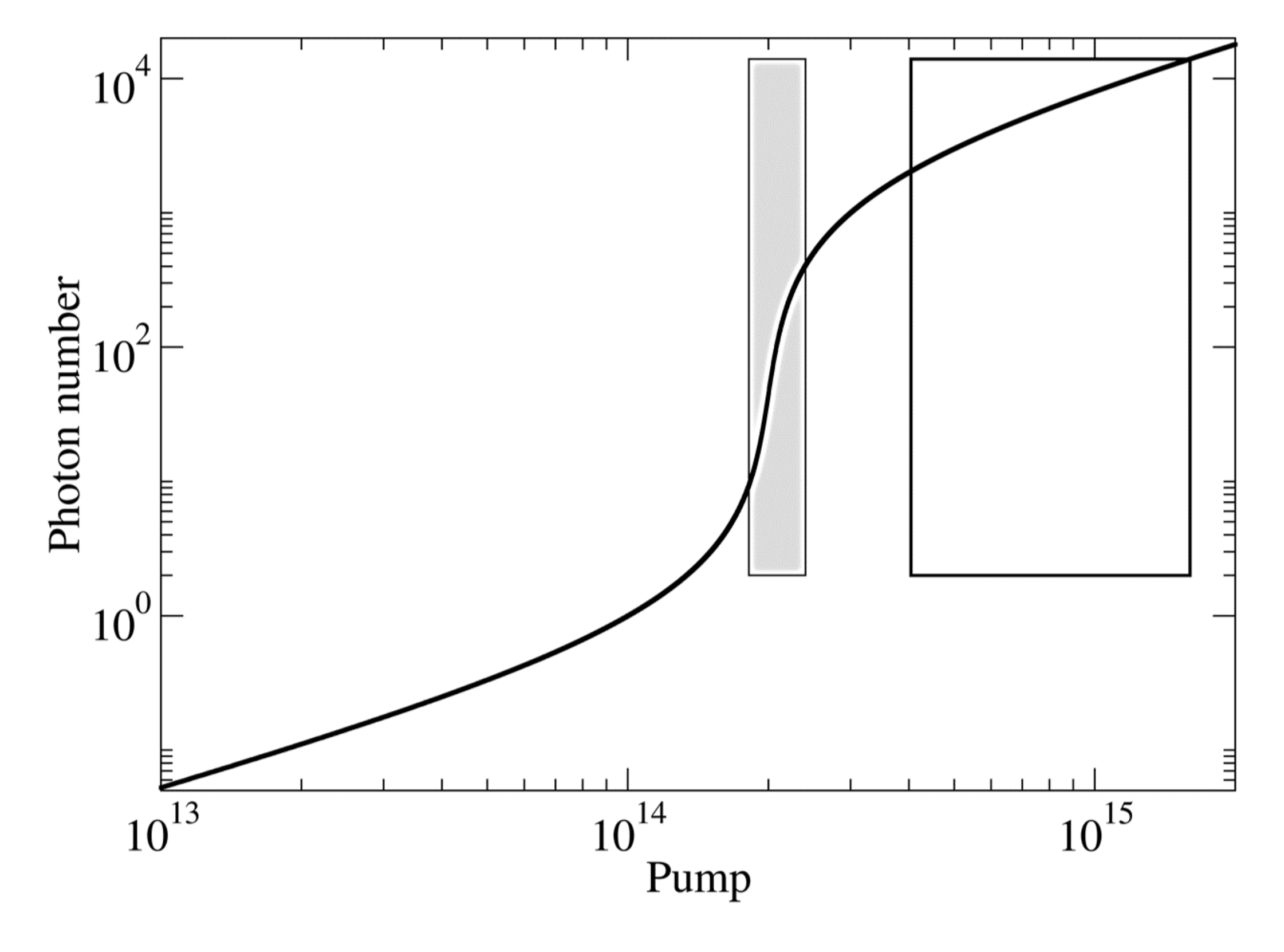}
  \caption{Modulation regions for pulse generation in the transition region, grey rectangle, and for data encoding above
threshold, empty rectangle, for laser response compatible with those of our device (L-L curve in Fig. \ref{ExperiIV}). The interval covered by the empty rectangle corresponds approximately to the current interval for large amplitude modulation of the
device used in this experiment ($i_{min} \approx 3 mA, i_{max} \approx 10 mA$)}
  \label{comp-mod}
\end{figure}

We now compare the results we have obtained to the standard above threshold modulation. Fig.\ref{comp-mod} illustrates qualitatively the two regimes, plotted as a grey rectangle superposed to the laser response for the modulation in the nonlinear transition region, as discussed previously, and the empty rectangle which qualitatively matches the modulation interval normally used in data encoding. The laser response is computed as in Fig. \ref{scurves}.

The standard estimate of the energy cost per pulse is based on the simple relation

\begin{equation}
\varepsilon_{p} = \frac{i_{b} \times V_{b}}{B} \, ,
\label{EC}
\end{equation}

\noindent where $V_{b}$ represents the bias voltage across the junction which matches the bias current, $i_{b}$ , and $B$ stands for the bit rate. This expression does not take into account the actual cost of the modulation and the nonlinear response due to the dynamic resistance.

Using eq. \ref{EC} we estimate for the conditions of Fig. \ref{dynamics}c - i.e., in the nonlinear transition regime - 

\begin{equation}
\varepsilon_{p,nl} \approx 2.3 pJ \times bit^{-1}\, ,
\end{equation}
\begin{equation}
\varepsilon_{opt} \approx 70 fJ \times bit^{-1}\, ,
\end{equation}

\noindent where we have taken $i_{b} = 1.15mA$, $V_{b} = 2V$ (average value from manufacturer’s datasheet ~\cite{Manufac}) and $B = 10^9 s^{-1}$ and the optical energy per bit, $\varepsilon_{opt}$, is estimated assuming a triangular pulse. $\varepsilon_{p,nl}$ represents the electrical energy cost per bit, while $\varepsilon_{opt}$ is the estimated optical energy per bit.

In order to make sense of these numbers one has to keep in mind that this is a low-cost (outdated) commercial device intended for low transmission rate ($2.5 Gbit/s$) in Gigabit Internet, rather than a top-of-the-line laser specially designed for low power consumption. Therefore we compare its performance obtained in the nonlinear response part of the curve to its standard, above threshold modulation.

Performing the same calculation for above threshold biasing (Non-Return-to-Zero) we obtain

\begin{equation}
\varepsilon_{p, at} \approx 5.2 pJ \times bit^{-1}\, ,
\end{equation}

\noindent where we have assumed $i_{b} = 6.5 mA$ for $V_{b} = 2V$ (as above) and $B = 2.5 \times 10^9 s^{-1}$ according to the Manufacturer's specifications ~\cite{Manufac}. Notice that we have implicitly assumed a square modulation with lower $i_{min} = 3 mA$ (to be compatible with the maximum threshold current specified by the manufacturer ~\cite{Manufac}) upper current level $i_{max} = 10mA$, thus attributing the mid-value to the bias current $i_{b}$ , to optimally exploit the optical contrast bewteen high and low light levels.

This comparison shows that in spite of the reduction in the bit rate - $1 Gbit/s$ compared to $2.5 Gbit/s$ for above-threshold modulation -, there is a gain of nearly a factor 2 in the electrical energy per bit in passing from the standard biasing to the nonlinear transition region, thus making the scheme already attractive. However, this estimate may still be too conservative, since eq. \ref{EC} does not take into account the cost of the modulation amplitude and the consequence of the dynamic resistance which adds to the electrical dissipation. As long as the modulation amplitudes and bias currents are the same,
there is a proportionality factor which one can assume to be nearly device independent, thus the information given by the simpler expression, eq.\ref{EC}, is sufficient, but the comparison between the two schemes requires a closer look.

A more complete expression for the energy cost per bit can be obtained as follows

\begin{equation}
E_{p} = \varepsilon_{p} + i_{mod}^2 \frac{dV}{dI}\, ,
\end{equation}

\noindent where $i_{mod}$ is the peak to peak modulation amplitude for the pumping current and $\frac{dV}{dI}$ the differential resistance specified by the manufacturer. With these values using $i_{mod,nl} = 0.6mA$ and $i_{mod,at} = 7mA$, with the maximum
value of the differential resistance specified for the device [22] $\frac{dV}{dI} = 65\Omega$, we obtain

\begin{equation}
E_{p,nl} \approx \varepsilon_{p,nl} = 2.3pJ \times bit^{-1}\, ,
\end{equation}
\begin{equation}
E_{p,at} \approx 6.5pJ \times bit^{-1}\, ,
\end{equation}

\noindent i.e., a negligible correction for the modulation in the transition region but an increase in the energy spent per
pulse in the traditional above threshold scheme, which results in an electrical dissipation approximately three times lower in the low biasing, low amplitude scheme. We can therefore conclude that there is a substantial gain in passing from the traditional above-threshold scheme to the generation of bits in the nonlinear region of the laser response.

One important point to remark upon is the fact that the transmission rate in this device is particularly low. Increasing it by an order of magnitude, with suitable device design, would bring the energy consumption in the range of $200 pJ /bit$ , which is not unlike the one of purpose-built devices. For DFB-based edge emitters, which require a much larger footprint than a VCSEL, the current record for power consumption is just under $100 pJ/bit$ ~\cite{Tomiyasu2018}. Since current VCSEL cavity designs are capable of transmitting at $35 Gb/s$ with an energy cost of $145 fJ/bit$ ~\cite{Li2015}, it is reasonable to expect, by extrapolation, that applying a modulation in the nonlinear response region should provide transmission rates exceeding $10 Gb/s$ with an energy cost of the order of $50 fJ/bit$, thus potentially enabling the technology to approach the needs of datacenters ~\cite{Cheng2018}.

\section{Conclusion}
The scheme we have presented is an exploratory investigation of the potential for exploiting the transition region between the two emission branches (spontaneous emission and stimulated light, respectively) to generate energy-efficient pulses for low power communications. The device we have used for the investigation is particularly flexible, thanks to its relatively low modulation frequency and the work is a proof of principle for the extension to faster, larger $\beta$ devices.

Measurements, supported by numerical modeling, have convincingly shown the existence of an optimal below-threshold time period ($\tau_{bt} \approx \gamma^{-1}$ ) which determines the modulation amplitude once the bias is chosen, depending on the amount of noise, while the optimal modulation frequency is determined on the basis of the device's spectral response [23].

Finally, we have shown how scaling the technique to top-of-the-line VCSELs promises to bring new records in energy-per-bit consumption at competitive transmission rates, where the pulses generated by this scheme would be a comb on which information would be encoded by removing the 0-bits.

\section*{Acknowledgment}

The authors are grateful to the R\'egion PACA and BBright for support, and to B. Garbin, F. Gustave and M. Marconi for assistance and discussions. Technical support from J.-C. Bery (mechanics) and from J.-C. Bernard and A. Dusaucy (electronics) is gratefully acknowledged. T. W. thanks the scientific research starting fund (KYS045618036) and NSFC (61804036).
)


\begin{thebibliography}{1}

\bibitem{Song2005}  
B. ~S. Song, S. ~Noda, T. ~Asano and Y. ~Akahane, ``Ultra-high-Q photonic double-heterostructure nanocavity,'' Nature Materials, \textbf{4}, 207-210, (2005).

\bibitem{Khajavikhan2012}
M. ~Khajavikhan, A. ~Simic, M. ~Katz, J. ~H. Lee, B. ~Slutsky, A. ~Mizrahi, V. ~Lomakin and Y. ~Fainman, ``Thresholdless nanoscale coaxial lasers,'' Nature, \textbf{482}, 204-207, (2012).

\bibitem{Hill2014}
M. ~T. Hill and M. ~C. Gather, ``Advances in small lasers,'' Nature Phot., \textbf{8}, 908-918, (2014).

\bibitem{Smit2012}
M. ~Smit, J. ~van der Tol, and M. ~Hill, ``Moore's law in photonics,'' Laser \& Photonics Rev., \textbf{6}, 1-13, (2012).

\bibitem{Soref2018}
R. ~Soref, ``Tutorial: Integrated-photonic switching structures,'' Appl. Phys. Lett. Phot., \textbf{3}, 021101, (2018).

\bibitem{Norman2018}
J. ~C. Norman, D. ~Hung, Y. ~Wan, and J. ~E. Bowers, ``Perspective: The future of quantum dot photonic integration,'' App. Phys. Lett. Phot., \textbf{3}, 030901, (2018).

\bibitem{Service2010}
R. ~F. Service, ``Nanolasers. Ever-smaller lasers pave the way for data highways made of light,'' Science, \textbf{328}, 810, (2010).

\bibitem{Chen2011}
R. ~Chen, T. ~T. ~D. Tran, K. ~W. Ng, W. ~S. Ko, L. ~C. Chuang, F. ~G. Sedgwick, and C. ~Chang-Hasnain, ``Nanolasers grown on silicon,'' Nature Phot., \textbf{5}, 170-175, (2011).

\bibitem{Huang2014}
K. ~C. Y. Huang, M. ~K. Seo, T. ~Sarmiento, Y. ~Huo, J. ~S. Harris, and M. ~L. Brongersma, ``Electrically driven subwavelength optical nanocircuits,'' Nature Phot., \textbf{8}, 244-249, (2014).

\bibitem{Liu2015}
N. ~Li, K. ~Liu, V. ~Sorger, and D. ~Sadana, ``Monolithic III-V on silicon plasmonic nanolaser structure for optical interconnects,'' Sci. Rep., \textbf{5}, 14067, (2015).

\bibitem{Willets2007}
K. ~A. Willets and R. ~P. Van Duyne, ``Localized surface plasmon resonance spectroscopy and sensing,'' Ann. Rev. Phys. Chem., \textbf{58}, 267-297, (2007).

\bibitem{Anker2008}
J. ~N. Anker, W. ~P. Hall, O. ~Lyandres, N. ~C. Shah, J. ~Zhao, and R. ~P. Van Duyne, ``Biosensing with plasmonic
nanosensors,'' Nature Mat., \textbf{7}, 442-453, (2008).

\bibitem{Gourley2008}
P. ~L. Gourley, D. ~Y. Sasaki, and R. ~K. Naviaux, ``Nanolaser spectroscopy for studying novel biomaterials,'' Proc. SPIE., \textbf{6859}, 685914, (2008).

\bibitem{Kim2013}
T. ~Kim, J. ~G. McCall, Y. ~H. Jung, X. ~Huang, E. ~R. Siuda, Y. ~Li, J. ~Z. Song, Y. ~M. Song, H. ~A. Pao, R. ~H. Kim, C. ~F. Lu, S. ~D. Lee, I. ~S. Song, G. ~C. Shin, R. ~A. Hasani, S. ~Kim, M. ~P. Tan, Y. ~G. Huang, F. ~G. Omenetto, J. ~A. Rogers, and M. ~R. Bruchas, ``Injectable, cellular-scale optoelectronics with applications for wireless optogenetics,'' Science, \textbf{340}, 211-216, (2013).

\bibitem{Lozovik2014}
Y. ~E. Lozovik, I. ~A. Nechepurenko, A. ~V. Dorofeenko, E. ~S. Andrianov, A. ~A. Pukhov, ``Spaser spectroscopy with subwavelength spatial resolution,'' Phys. Lett. A, \textbf{378}, 723-727, (2014).

\bibitem{Abe2015}
H. ~Abe, M. ~Narimatsu, T. ~Watanabe, T. ~Furumoto, Y. ~Yokouchi, Y. ~Nishijima, S. ~Kita, A. ~Tomitaka, S. ~Ota, Y. ~Takemura, and T. ~Baba, ``Living-cell imaging using a photonic crystal nanolaser array,'' Opt. Express, \textbf{23}, 17056-17066, (2015).

\bibitem{Editorial2017}
Editorial, ``Scrutinizing Lasers,'' Nature Photonics, \textbf{11}, 139, (2017).

\bibitem{Rice1994}
P. ~R. Rice and H. ~J. Carmichael, ``Photon statistics of a cavity QED laser: A comment on the laser phase transition analogy,'' Phys. Rev. A, \textbf{50}, 4318-4329, (1994).

\bibitem{Wang2018}
T. ~Wang, G. ~P. Puccioni, and G. ~L. Lippi, ``Threshold dynamics in meso- and nanoscale lasers: why Vertical Cavity Surface Emitting Lasers?,'' Proc. SPIE, \textbf{10682}, 106820Q, (2018).

\bibitem{Khurgin2012}
J. ~B. Khurgin and G. ~Sun, ``How small can Nano be in a Nanolaser?,'' Nanophotonics, \textbf{1}, 3-8, (2012).

\bibitem{Wang2015}
T. Wang, G. P. Puccioni and G. L. Lippi, ``Dynamical Buildup of Lasing in Mesoscale Devices,'' Sci. Rep., \textbf{5}, 15858, (2015).

\bibitem{Manufac}
\url{Manufacturer's specifications: https:// www.thorlabs.de/drawings/d6e6bf60a1b18408-F0BC92E1-BA82-C641-BB9776F0E1974E15/VCSEL-980-MFGSpec.pdf/}

\bibitem{Wang2018a}
T. ~Wang, G. ~P. Puccioni, and G. ~L. Lippi, ``Onset of lasing in small devices: the identification of the first threshold
through autocorrelation resonance,'' Ann. Physik, \textbf{1800086}, 1-7, (2018).

\bibitem{Porta1998}
P. ~A. Porta, L. ~M. Hoffer, H. ~Grassi, and G. ~L. Lippi, ``Control of turn-on in Class B lasers,'' Int. J. Bifurcations
Chaos Appl. Sci. Eng., \textbf{8}, 1811-1819, (1998).

\bibitem{Porta2000}
P. ~A. Porta, L. ~M. Hoffer, H. ~Grassi, and G. ~L. Lippi, ``Analysis of a nonfeedback technique for transient steering
in class-B lasers,'' Phys. Rev. A, \textbf{61}, 033801, (2000).

\bibitem{Arecchi1984}
F. ~T. Arecchi, G. ~L. Lippi, G. ~P. Puccioni, and J. ~R. Tredicce, ``Deterministic chaos in laser with injected signal,'' Opt. Commun., \textbf{51}, 308-314, (1984).

\bibitem{Tredicce1985}
J. ~R. Tredicce, F. ~T. Arecchi, G. ~L. Lippi, and G. ~P. Puccioni, ``Instabilities in lasers with an injected signal,'' J. Opt. Soc. Am. B, \textbf{2}, 173-183, (1985).

\bibitem{Puccioni2015}
G. ~P. Puccioni and G. ~L. Lippi, ``Stochastic Simulator for modeling the transition to lasing,''Opt. Express, \textbf{23}, 2369-2374, (2015).

\bibitem{Einstein1917}
A. Einstein, ``Zur Quantentheorie der Strahlung,'' \textit{Physikalische Zeitschr}, \textbf{18}, 121-128, (1917).

\bibitem{Tomiyasu2018}
T. ~Tomiyasu, D. ~Inoue, T. ~Hiratani, K. ~Fukuda, N. ~Nakamura, T. ~Uryu, T. ~Amemiya, N. ~Nishiyama, and S.~Arai, ``20-Gbit/s direct modulation of GaInAsP/InP membrane distributed-reflector laser with energy cost of less than 100 fJ/bit,'' Appl. Phys. Express, \textbf{11}, 012704, (2018).

\bibitem{Li2015}
H. ~Li, P. ~Wolf, P. ~Moser, G. ~Larisch, J. ~A. Lott,and D. ~Bimberg, ``Temperature-Stable, Energy-Efficient, and High-Bit Rate Oxide-Confined 980-nm VCSELs for Optical Interconnects,'' IEEE J. Sel. Topics Quantum Electron., \textbf{21}, 405-413, (2015).

\bibitem{Cheng2018}
Q. Cheng, S. Rumley, M. Bahadori, and K. Bergman, ``Photonic switching in high performance datacenters,'' Opt. Express, \textbf{26}, 16022-16043, (2018).

\end{thebibliography}
\end{document}